\documentclass{article}

\usepackage{PRIMEarxiv}

\usepackage[utf8]{inputenc} % allow utf-8 input
\usepackage[T1]{fontenc}    % use 8-bit T1 fonts
\usepackage{hyperref}       % hyperlinks
\usepackage{url}            % simple URL typesetting
\usepackage{booktabs}       % professional-quality tables
\usepackage{amsfonts}       % blackboard math symbols
\usepackage{nicefrac}       % compact symbols for 1/2, etc.
\usepackage{microtype}      % microtypography
\usepackage{lipsum}
\usepackage{fancyhdr}       % header
\usepackage{graphicx}       % graphics
\graphicspath{{media/}}     % organize your images and other figures under media/ folder

\title{Opinion polarization in human communities can emerge as a natural consequence of beliefs being interrelated
\thanks{\textit{\underline{Citation}}: 
\textbf{Anna Zafeiris, "Opinion polarization in human communities can emerge as a natural consequence of beliefs being interrelated",  Entropy 2022, 24(9), 1320; https://doi.org/10.3390/e24091320}} 
}

\author{
  Anna Zafeiris\\
  MTA-ELTE 'Lend\"{u}let' Collective Behaviour Research Group\\
  Hungarian Academy of Sciences, E\"{o}tv\"{o}s~University, 1117 Budapest, Hungary\\
  \texttt{\ anna.kinga.zafeiris@ttk.elte.hu} }

\begin{document}
\maketitle

\begin{abstract}
The emergence of opinion polarization within human communities---the phenomenon that individuals within a society tend to develop conflicting attitudes related to the greatest diversity of topics---has been a focus of interest for decades, both from theoretical and modelling points of view. Regarding modelling attempts, an entire scientific field---opinion dynamics---has emerged in order to study this and related phenomena. Within this framework, agents' opinions are usually represented by a scalar value which undergoes modification due to interaction with other agents. Under certain conditions, these models are able to reproduce polarization---a state increasingly familiar to our everyday experience. In the present paper, an alternative explanation is suggested along with its corresponding model. More specifically, we demonstrate that by incorporating the following two well-known human characteristics into the representation of agents: (1) in the human brain beliefs are interconnected, and (2) people strive to maintain a coherent belief system; polarization immediately occurs under exposure to news and information. Furthermore, the model accounts for the proliferation of fake news, and shows how opinion polarization is related to various cognitive biases.
\end{abstract}

% keywords can be removed

\section{Introduction}

As Evan Williams, co-founder of Twitter, famously said in a 2017 interview, ``I thought once everybody could speak freely and exchange information and ideas, the~world is automatically going to be a better place. I was wrong about that''~\cite{TwitterIiB}. This is a good overview of the surprise that the unprecedented connectivity among people---primarily driven by various internet-based social media platforms in the early 21{st} century---brought us unprecedented factions, dissension and fake news~\cite{MisDisInfos, SajatFN}, instead of agreement and conciliation~\cite{DividedB1, DividedB2}. The~reasons behind these phenomena are diverse and manifold and, accordingly, are the subject of the most diverse scientific fields from history~\cite{FergusonNW} through sociology~\cite{HomophSoc} to computational social science~\cite{Dandekar5791, PerraSciRep, SajatSciRep}.

The quest for finding a valid explanation---and a practicable model---for the phenomena of the above mentioned polarization and fragmentation has been underway for decades. Specifically, regarding computational models, an~entire field, \emph{opinion dynamics} %MDPI: Please confirm in the whole paper if the italics should be retained. Answer: Thank you. They should be.
, has emerged in order to study the way opinions, information, views and beliefs propagate in human communities~\cite{OpDynOf2020, ContOpDynSurvey, HegselmannKrause02, Olle14, OrigDeffuant, Rev3_C}. In~general, these models study the dynamics of attitudes related to a certain topic, such as issues related to climate change, abortion, immigration, vaccination, a~certain politician or political party, etc. Typically, the~attitudes of the agents towards the given issue are described by scalar values which are assumed to be altered due to communication with peers~\cite{OpDynOf2020, ContOpDynSurvey, AlapOpDynOf}. Within~this scientific field, \emph{consensus} refers to the state in which all social actors share the same opinion, \emph{polarization} to the condition when each agent accepts one of two opposing opinions, while \emph{fragmentation} refers to a transitory phase, in~which a finite number of distinct ``opinion islands''~appear.

A key concept related to continuous opinion dynamics models---models representing opinions as scalars taken from a continuous interval usually between $-1$ and $+1$---is the ``confidence threshold'', which is a value above which agents cease to communicate with each other. Above~this threshold, ``classical'' bounded confidence models predict consensus, and~fragmentation and/or polarization under it~\cite{OrigDeffuant, Olle14, HegselmannKrause02, AlapOpDynOf}. However, the~polarization/fragmentation breaks down in the presence of noise~\cite{HianyoltCikk1, HianyoltCikk2, HianyoltCikk3, HianyoltCikk4, LiuSciRep13}---an essential and unavoidable component in all social and biological systems~\cite{AlapOpDynOf}. In~order to reach polarization, other mechanisms have been suggested, such as ``distancing'', which is the direct amplification of differences between dissimilar individuals~\cite{HianyoltCikk1}. Lately, these models have been further developed by incorporating more realistic features, such as heterogeneity with respect to the agents' confidence thresholds~\cite{UjRef1Realstc1} or by incorporating an underlying communication network~\cite{UjRef1Realstc2} defining the pattern by which social actors interact with each other. These approaches lead to more complex---and realistic---dynamics. Furthermore, recently, building on the observation that \emph{events} often have a polarization effect~\cite{Event01, Event02}, the~process of polarization was modelled by extending a classical bounded confidence model---the so called Hegselmann-Krause model~\cite{HegselmannKrause02}---in a way that individuals change their opinions in line with the certain event~\cite{Event03}. Other models approach the problem from a kinetic~\cite{Rev3_B} or hydrodynamic~\cite{Rev3_A} perspective.

In the present paper, an~alternative, or~complementary, explanation is suggested, by~showing that in case some ``basic'' human characteristics are incorporated, polarization immediately appears once agents are exposed to new information --- even without direct communication. These almost trivial human characteristics are that (i) human beliefs are interrelated rather then evolving independently of each other, and~(ii) people strive to maintain a coherent, contradiction-free belief-system. 
The fundamental difference between "classical" opinion dynamics models and the ones incorporating such features is that while in the first case agents are represented with a single scalar value, in~the second case the model of the social actors has some kind of inner structure. This difference gives rise to an entirely different dynamic~\cite{Maas1, Maas2}.

%%%%%%%%%%%%%%%%%%%%%%%%%%%%%%%%%%%%%%%%%%

\section{The Background: Fundamental Features of Human Belief~Systems}\label{sec:backgr}

During the last decades, a~vast amount of knowledge has accumulated in scientific fields on the ways humans perceive the world, make decisions and structure their beliefs. Despite the fact that obtaining a detailed understanding of these processes still require further work, some scientific fields, such as neurobiology~\cite{EBGoldsteinCognPsy, Seth, Pinker, CogPsyTK, KCC} or various human sciences~\cite{Sapiens, Searle10}, such as psychology~\cite{BelBrain, CognIll, KnowldgIllsn, CDL:Cooper07}, anthropology~\cite{CsanyiHied}, economics~\cite{KahnemanTFS, MarchDM, IdentityEcon} and political science~\mbox{\cite{Converse1964, JostFedNap2009, OilSpill, ConservVsLib, Rokeach63, DividedB1, DividedB2}} have progressed considerably. From~our point of view, the~key finding is that in humans, opinions and beliefs \emph{never} occur alone, that is, no concept or belief can exist in isolation. (Actually, humans are not even able to memorize anything without connecting it to something meaningful~\cite{EBGoldsteinCognPsy}.) Rather, concepts and beliefs are organized into a structure, a~system (\emph{"belief system"}) which has well-defined features~\cite{EBGoldsteinCognPsy, Converse1964, JostFedNap2009, GoldbergStein18}:

\begin{enumerate}
  \item \emph{First} and fore-most, we seek to be \emph{consistent}. This means that people try to maintain a belief system in which the elements mutually support each other, or~are independent~\cite{JostFedNap2009, GoldbergStein18}. In~the case of holding \emph{conflicting} beliefs, people experience discomfort called \emph{cognitive dissonance}~\cite{CDL:Festinger57, CDL:MedicalNewsT} which people will try to reduce. By this time, cognitive dissonance has become one of the most influential and well-researched theories in social psychology~\cite{CognDissA, CDL:MedicalNewsT, CDL:FrontiersInPsy13}.
   
  \item \emph{Secondly}, beliefs are not equally important: those that are more personal, closer to the ``self'', ``identity'' or ``ego'', trigger more intense feelings and are more difficult to change~\cite{CDL:MedicalNewsT, CsanyiHied, IdentityEcon, Scout}. In~this sense, beliefs have a \emph{hierarchical property}: the ones that are higher in rank define or constrain the ones that are lower in rank~\cite{SajatHierK}. For~example, the~belief or disbelief in God is a central (high-ranking) element in one's belief system, while the belief that ``an egg should be boiled for seven minutes in order to get the best soft-boiled egg'' is a low ranking one, and, accordingly, can be changed more easily~\cite{CsanyiHied}. Furthermore, beliefs that people hold in high regard tend to cause greater dissonance in case of contradiction with other beliefs~\cite{CDL:MedicalNewsT}.

  \item \emph{Finally}, beliefs belonging to the same broader topic (e.g., health, art-related topics, religion, political issues, etc.) are more strongly interrelated than beliefs belonging to different topics. For~example, attitudes towards ``freedom of speech'', ``religious freedom'' and ``freedom to choose spouse'' are more closely related than beliefs regarding ``freedom of speech'' and, say, homeopathic treatments. In~mathematical (graph-theoretical) terms, belief systems are ``modular'' or ``compartmentalized''~\mbox{\cite{CsanyiHied, ConservVsLib}.}

\end{enumerate}

\begin{figure}[ht]
\centering
\includegraphics[width=\linewidth]{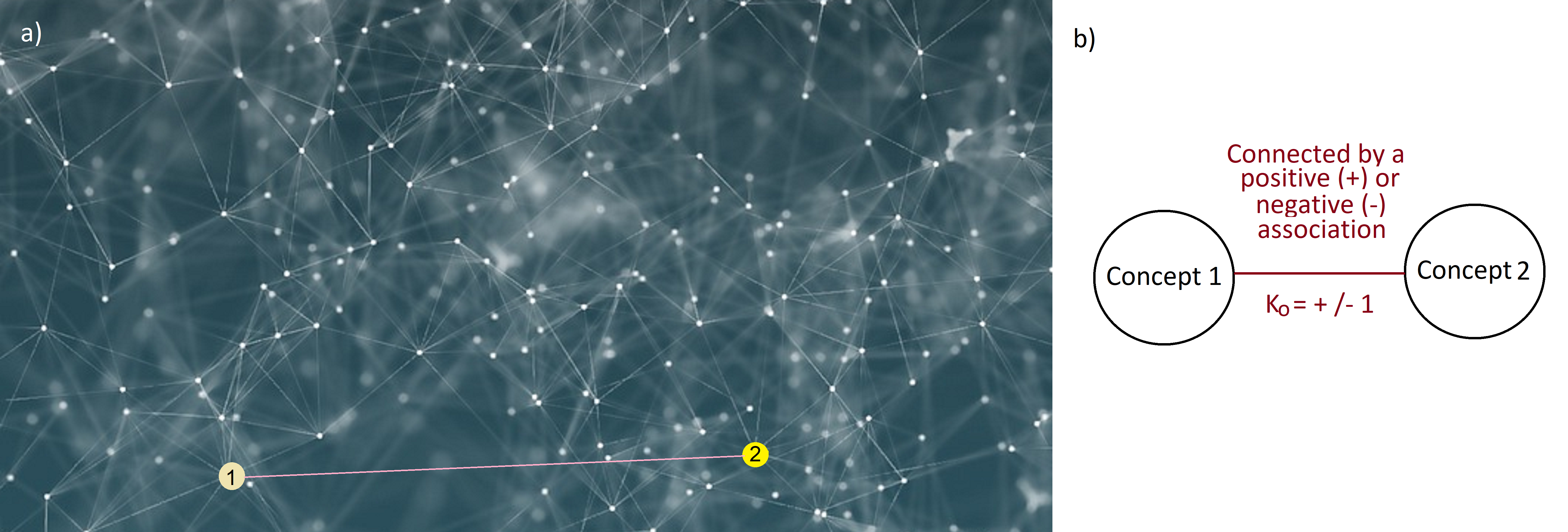}
\caption{The graph representation of belief systems. (\textbf{a}) As a first approximation, human belief systems can be represented by networks in which nodes are beliefs (``elements of a belief system''~\cite{EvalPolBSNW}) and edges represent relationships. (\textbf{b}) Links (relationships) can be either supportive (positive) or contradictory (negative)~\cite{Rodriguez16}.}  \label{fig:BSNW}
\end{figure}

As a first approximation, such a structure can be represented as a \emph{modular}, \emph{hierarchical network} (graph) in which the nodes are connected by supportive (positive) or invalidating (negative) relations~\cite{JostFedNap2009, OilSpill, EvalPolBSNW, ConservVsLib} (Figure~\ref{fig:BSNW}). For~example, according to a widespread belief, ``Pathogens can cause diseases''. According to another wide-spread belief---primarily in historical societies---``Diseases are caused by evil spirits''~\cite{EvolOfGod, DCh14}. These two concepts are in negative relation: somebody believing in one of these opinions will probably disagree with the other. In~contrast, the~beliefs ``Pathogens can cause diseases'' and ``Contagion is due to the spread of pathogens'' support each other (positive relationship), since accepting one of them renders the acceptance of the other more~probable.

This last requirement ensures that the \emph{level of consistency} can be defined~\cite{Rodriguez16}. In~this description, nodes are beliefs, and~edges represent functional relationships between \linebreak them~\mbox{\cite{BaldGold14, EvalPolBSNW, GoldbergStein18}.} This approach has already been applied by sociologists and economists as well in order to model political belief system dynamics~\cite{OilSpill, ConservVsLib, EvalPolBSNW, BaldGold14, PolBSNWCentral}. Within~this framework, the~focus is on the \emph{relatedness} of beliefs---captured by graph representation---while the hierarchical and modular characteristics do not gain special importance (see the Supplementary Material for the results incorporating the hierarchical characteristics as~well).

In these models, a~node is ``an element of a person's belief system''~\cite{EvalPolBSNW}, which the related literature names slightly differently, such as ``opinions''~\cite{OilSpill}, ``concepts''~\cite{Rodriguez16} ``attitudes''~\cite{OilSpill, EvalPolBSNW, PolBSNWCentral}, ``beliefs''~\cite{Rodriguez16, ConservVsLib, EvalPolBSNW}, or~``positions''~\cite{BaldGold14, PolBSNWCentral}. In~the present article,  the~terms \emph{``belief''}, \emph{``attitude''}  and \emph{``concept''} are mostly used, under~the condition that we consider all sorts of human thoughts as an ``element of a belief system'' (that is: a ``belief''), whether they be simple or complex, that can be transmitted with the use of language from one person's mind to another's~\cite{Searle10, CsanyiHied}. Accordingly, a~node can be a thought or attitude towards public issues such as abortion, climate change, immigration, vaccination, gun control; it can be information regarding a public figure or a political party, an~idea related to ``proper behaviour'', ``justice''; or the belief that somebody did or said something, etc. Furthermore, this definition implies that beliefs can be transmitted via communication. Communication, in~its most basic form, can be discussion (talking) between two or more individuals, but~it can also be news/beliefs/information spread by a single agent or organization to many people at the same time, for~example via public and social media, news channels, journals, etc. In~short, communication is the circulation of news, information and beliefs within a certain~community.

The fact that beliefs (attitudes/concepts) are in functional relation with each other (that is, if~they are connected, they either support or contradict each other) is crucial, because~it implies that some ``new'' belief will fit into an already existing system---the ones that \emph{in}crease the system's consistency---while others---the ones \emph{de}creasing the system's consistency---will not~\cite{GoldbergStein18}. These latter gives rise to the disturbing feeling of cognitive dissonance and people will apply various strategies in order to avoid them. Among~other strategies, they will try to keep contact only with those from whom they expect reassuring information (homophily), they will ignore certain information and focus on other information (attentional bias), while greater credence will be given to evidence that fits with the existing beliefs (confirmation bias). These strategies are known as various \emph{biases} in the field of psychology and sociology~\cite{CognIll, Bias2, Bias3, Dobelli14}. Different people apply different strategies to various extents; however, to~some level, all the strategies are applied by all of us~\cite{CognIll, KnowldgIllsn, Bias2, Bias3}.

Furthermore, these cognitive dissonance avoiding mechanisms are in close relation with the proliferation of fake news and the circulation of various types of questionable information as well. In~case the well-fitting of a piece of information into the already existing belief system weights more than its credibility, people will adopt it---simply because it provides the pleasant feeling of reassurance. This mechanism is applied in all aspects of life, not only in case of political issues. For~example, in~the field of economics, it has been observed that managers, whose sales fall short of expectations, rather than rethinking the qualities of the product, tend to identify the cause of the failure elsewhere, for~example in the marketing campaign. In~such cases, they state that the \emph{marketing} campaign failed, so it is actually a miracle that the product was sold at all~\cite{GeorgeJones2001}. By~finding this new explanation, the~sales results show directly the \emph{merits} of the product, not its failure, and~as such, serves as a basis for the pleasant feeling of reassurance. This mechanism is analyzed in Section~\ref{sec:NewBelfResult}.

%%%%%%%%%%%%%%%%%%%%%%%%%%%%%%%%%%%%%%%%%%
\section{The~Model}\label{sec:Model}

We assume a population of $N$ agents. At~this point, only their attitudes towards two concepts are important (the attitudes towards, say, vaccination (Concept 1) and, say, a~certain public figure (Concept 2)). People can hold any kind of attitudes towards these concepts, from~total condemnation (marked by $-1$) to total support (denoted by $+1$). Neutrality or indifference is indicated by zero or near-zero values. We are interested in how the agents' attitudes evolve due to being exposed to some news (piece of information) that creates a relation between two originally independent concepts (see Figure~\ref{fig:BSNW}(b). 

The relation $K_0$ can be positive or negative. Using the above example, a~trivial positive connection can be that ``XY public figure (concept 2) has spoken out \emph{in favor} of vaccination (concept 1)'' ($K_0=+1$), while a negative connection can be that ``XY public figure has spoken out \emph{against} it'' ($K_0=-1$). 

In case an agent holds positive attitudes towards both concepts, the~positive message (support of vaccination) will give rise to the comforting feeling of \emph{reassurance}~\cite{KnowldgIllsn}. In~this case, the~original attitudes are reinforced, since both concepts become better connected and further embedded into the belief system. In~contrast, in~case the XY trusted and respected politician takes a position \emph{against} vaccination, a~supported matter, the~agent will experience \emph{cognitive dissonance} with an intensity proportional to the original attitude values, and~will apply a strategy in order to reduce it~\cite{CDL:MedicalNewsT, CDL:Festinger57}. 

Turning back to the basic scenarios, an~agent can hold negative attitude towards one of the concepts, and~a positive one towards the other---say, for~example, a~negative attitude towards the public figure and positive attitude towards vaccination. In~this case, a~negative relation will give rise to reassurance,i.e., ``XY politician, whom I anyway hold very low, talked out against vaccination, a~cause so important for me ... No surprise here, a~fool is known by his conversation'' and so on. All scenarios can be analyzed with the same train of~thought. 

Accordingly, from~a mathematical point of view, the~cognitive dissonance (or reassurance), ${C_i(t)}$, that agent $i$ will experience at time-step $t$, can be formulated as:
\begin{equation}\label{eq:C}
C_i(t) = a_{i,1}(t) \cdot a_{i,2}(t) \cdot K_0 \
\end{equation}

\noindent where $a_{i,1}(t)$ and $a_{i,2}(t)$ are the original attitudes of agent $i$ towards concept 1 and 2, respectively, at~time-step $t$ , and~$K_0$ is the type of connection, which can take two values, $+1$ or $-1$, according to the supportive or opposing nature of the connection between the concepts (see also Figure~\ref{fig:BSNW}(b). In~case $C$ is positive, it is called \emph{reassurance}, while in case it is negative, it is usually referred to as \emph{cognitive dissonance}. Anyhow, in~both cases, $C$ denotes the value by which the information alters the \emph{coherence} or \emph{consistency} level of agent $i$'s belief system. Note that $\mid C_i(t) \mid \leq 1$ is always the case, since $\mid a_{i,1}(t)\mid \leq 1$, $\mid a_{i,2}(t)\mid \leq 1$, and ~$K_0 = \pm 1$.%MDPI: we changed +/- to plus minus sign, please confirm. CONFIRMED

According to the literature~\cite{CognIll, KnowldgIllsn, Bias2, Bias3}, in~case of facing information inducing cognitive dissonance, people attempt to relieve the discomfort in different ways, among~which the most common ones are:

\begin{enumerate}
\item[(i)] Rejecting new information that conflicts with the already existing ones;
\item[(ii)] Re-evaluating the attitudes;
\item[(iii)] A tendency of ``explaining things away'', that is, finding alternative explanations (developing new beliefs) which supplement the original information in a way that the primordial contradiction is~dissolved.
\end{enumerate}

From a modeling point of view, the~first strategy---rejecting the information---simply leaves the belief system unaltered. In~this case, in~the framework of the model, the~network---nodes, edges and weights---remain unchanged. The~second and third strategies do modify the belief system, due to the new connection between the originally unconnected concepts. In~the following section, we will focus on modelling these~strategies.

\subsection{Modelling the Re-evaluation of~Beliefs}\label{sec:ReevalBSModel}

The constant re-evaluation of our already existing beliefs is an inevitable part of the process of learning and development~\cite{EBGoldsteinCognPsy}. New information often comes in the form of creating connection among concepts and beliefs that were originally disconnected. As~a matter of fact, this is a basic form of learning. Furthermore, people tend to evaluate most information, beliefs and concepts according to some personal narrative, a~personal ``frame of mind'', which is different from person to person. Simply put, this unique narrative is our personality~\cite{Sapiens}, which defines the very way we perceive the world and make decisions~\cite{IdentityEcon, MarchDM}. This variety entails individual differences in evaluating the most diverse topics around us, whether it be the judgement of a public figure, a~movie or the question of~immigration.

In the context of a formal model, the~most simple and plausible way to grasp these attitudes is to use numbers between $-1$ and $+1$ in a way that negative values represent negative attitudes and positive ones refer to positive stances. The~two extreme values, $-1$ and $+1$, refer to complete condemnation/approval, respectively.

In order to see \emph{how} these values might change, consider for example the following case: Paul believes that, say, genetically modified food is harmful. He has already heard it from his friends, and~now he reads it in his favorite blog as well. This gives him a feeling of reassurance, due to which he will be convinced about the verity of this belief even more, and~will be more attached to his favorite blog as well. In~other words, the~"embeddedness" of the original attitudes will increase. Mathematically speaking, his already positive attitudes towards these concepts (his belief and the blog) will increase even more due to the positive connection. Now consider a situation where he learns the opposite from his favorite blog, namely that there is nothing at all that could be harmful in genetically modified food (that is, a~\emph{negative} association appears among the two positive concepts: the belief and the blog). In~this case, he will experience some level of cognitive dissonance, whose extent depends on his original commitments towards the two concepts~\cite{CDL:Festinger57, CDL:Cooper07, CDL:FrontiersInPsy13}. This experience will make him less convinced, either of the reliability of the blog or of the belief itself---or both. Mathematically speaking, the~originally positive values (attached to the two concepts) will decrease somewhat. In~other words, \emph{cognitive dissonance} (negative $C_i(t)$ values) \emph{de}creases the absolute value of the affected attitude ($k$), while \emph{reassurance} (positive $C$ values) \emph{in}creases it. Consider the following formula:
\begin{equation}\label{eq:att}
a_{i,k}(t+1) = sign(a_{i,k}(t)) \cdot ( \mid a_{i,k}(t) \mid + \rho \cdot C_i(t) ) + Z_A
\end{equation}

\noindent where $a_{i,k}(t)$ is the original attitude of agent $i$ at time-step $t$ towards attitude $k$, $sign(a_{i,k}(t))$ is its signal ($+$ or $-1$), $\rho$ is a random value ("noise") taken from the $[0, 1]$ interval with uniform distribution, effecting the extent to which the attitude changes,  and~$C_i(t)$ (defined by Equation~(\ref{eq:C})), is the level of "coherence" (commonly known as \emph{cognitive dissonance}, in~case it is negative, and~\emph{reassurance} in case it is positive). 
Finally, the~$Z_A$ noise comprises the effects of other factors influencing the change of attitudes. It can be either positive or negative with equal probability. In~case the updated attitude value $a_{i,k}(t+1)$ falls outside the predefined $[-1, 1]$ interval, it is set to the nearest threshold ($+1$ or$-1$).

Attitudes do not vary with the same probability and to the same extent in case of different people and topics; for some, environmental issues are extremely important (and ``nothing can change'' this stance), some people are detached, while others are convinced that they are just evil-minded hoaxes. The~more extreme an attitude is, the~more difficult is to change it~\cite{CDL:MedicalNewsT, CsanyiHied, IdentityEcon, Scout}. (See also Section~\ref{sec:backgr}, 2{nd} bulleted point, "hierarchical property" of belief systems).

Mathematically speaking, the~feature \emph{"more difficult to change"} can be introduced into the model in two~ways:
\begin{enumerate}
    \item The more extreme an attitude value $a$ is, the~lower the \emph{probability} that it will change. Equation~(\ref{eq:AttChProb}) expresses the most simple mathematical formulation of this relation.
    \item The more extreme an attitude value $a$ is, the~smaller the \emph{magnitude} with which it can change. 
\end{enumerate}

For the results presented in the main text, the~above mentioned hierarchical property was introduced into the model according to the first way, that is, by~setting the \emph{probability} $p(AttCh_{i,k}(t))$ of attitude-change according to Equation~(\ref{eq:AttChProb}), and~setting $\rho$---the parameter controlling the maximal \emph{extent} with which the attitude values alter due to the experienced cognitive dissonance or reassurance ($C$)---to 1. In~other words, in~Equation~(\ref{eq:att}), $\rho = 1$ for the results presented in the main text. In~the Supplementary Material, a~detailed analysis is provided on how the parameter $~\rho$ effects the simulations (leading to the conclusions that the main claims remain valid, independently of the maximal extent of the alterations, see Figure~S7).
\begin{equation}\label{eq:AttChProb}
p(AttCh_{i,k}(t)) = 1 - |a_{i,k}(t)|
\end{equation}

Note that in all the equations, the~updated attitude values depend only on the agents' previous attitude values ($a_{i,k}(t)$), the~type of the news ($K_0$), and~on the cognitive dissonance (or reassurance) values that the news creates in the agents ($C_i(t)$). This means that agents develop their attitudes independently from each other. This originates from the fact that \emph{the source of the information does not matter} in the present model. Accordingly, if~the assumption is that it is the \emph{agents} who circulate the news among themselves (for example in the form of ``gossiping'' either in person or on social media), then they interact with each other. In~contrast, if~the source of the information is something else (for example, state media or some kind of propaganda) then agents do not interact directly with each other. In~reality, information usually circulates in both ways.
The reason why entire populations are considered is twofold. Firstly, because~one single agent cannot \emph{``polarize''}; they can develop extreme attitudes under certain circumstances. \emph{Polarization} is an emergent, statistical property of communities, a~phenomenon which does not have an interpretation on the level of individuals. The~larger the statistics, the~more apparent the phenomenon. The~second reason is that after studying the elementary process of attitude-update in detail (which is the topic of the present paper), an~immediate next step is to study the way by which agents manipulate and organize their social ties (links) assuming similar motivations (avoiding cognitive dissonance and enjoying reassurance).

\subsection{Modelling the Inclusion of New Beliefs in Order to Relieve Cognitive~Dissonance}\label{sec:NewBelInclModel}

In case a social actor experiences the upsetting feeling of cognitive dissonance due to a certain piece of information, a~commonly applied strategy is to adopt---or create---an even newer belief that changes the context of the original one in a way that it does not serve as a basis of cognitive dissonance any longer; rather, it becomes neutral or even gives rise to the pleasant feeling of reassurance~\cite{GeorgeJones2001, KnowldgIllsn, CognIll}. An~example of this maneuver is mentioned at the end of Section~\ref{sec:backgr}, related to managers whose sales data lag behind the expectations tendentiously conceive of various explanations, e.g,. ones related to ``awfully managed'' marketing campaigns. By~adopting this new belief (namely that the marketing campaign was awfully managed), the~cognitive dissonance caused by the negative sales results (linking a failure to their ``self'') is eliminated; furthermore, in~this light, the~sales-results could be seen as an achievement rather than a~failure.

The most simple assumption is that the probability $p(NB_i(t))$ of adopting such a new belief (by agent $i$ at time-step $t$) is proportional to the relief its adoption provides. Since only positive $C^{NB}_i(t)$ values represent reassurance, the~most simple mathematical formula is the following:
\begin{equation}\label{eq:NewBelAdoptProb}
p(NB_i(t)) = max(0, C^{NB}_i(t))
\end{equation}

where,
\begin{equation}\label{eq:CNB}
C^{NB}_i(t) = a_{i,1}(t) \cdot a_{i,3}(t) \cdot K_{NB} \
\end{equation}
 As shown before, $a_{i,1}(t)$ and $a_{i,3}(t)$ are the attitudes of agent $i$ towards concept 1 and the new belief at time-step $t$, respectively, and~$K_{NB}$ is the (positive or negative) connection type between them. Note that in case $C^{NB}_i(t)$ is negative---marking cognitive dissonance, instead of reassurance---the agent is highly unlikely to adopt the new~belief.

\section{Results}

\subsection{Re-Evaluating~Beliefs}\label{sec:ReevalBSResults}

Let us consider a population in which the agents' initial attitudes towards two arbitrarily chosen concepts are distributed uniformly, taking values from the $[-1, 1]$ interval. In~other words, at~the beginning of the simulation, all sorts of attitudes are present in the population with equal probability, from~complete condemnation to complete support and everything in between, with~an average of zero. Let us now assume that this population is exposed to some kind of news, connecting the two originally unconnected~concepts.

\begin{figure}[ht]
\includegraphics[width=\linewidth]{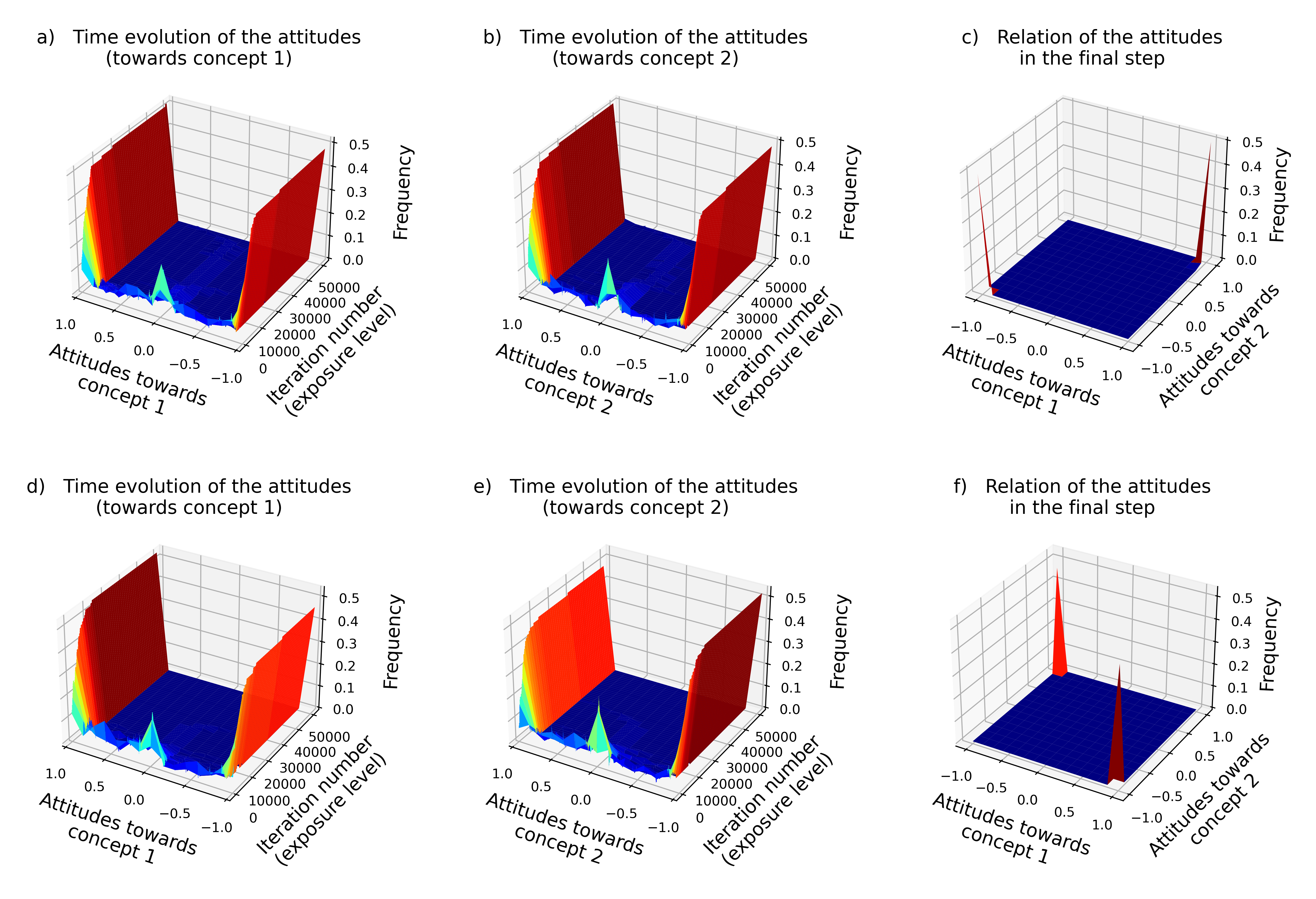}
\caption{A typical time evolution of two attitudes ($a_{i,1}(t)$ and $a_{i,2}(t)$) within a population of $N=100$ agents. Top row: $K_0 = +1$ (supportive relation). (\textbf{a},\textbf{b}): the distribution of attitudes values towards concepts 1 and 2, respectively, as~a function of time $t$. (\textbf{c}): At the final state, the~vast majority either supports both beliefs (marked by the peak at $(1, 1)$) or rejects it (marked by the peak at $(-1, -1)$). Bottom row (\textbf{d},\textbf{e}, and \textbf{f}): $K_0=-1$ (conflicting relation). (\textbf{f}) The major difference in this case is that at the end of the simulation most agents support one of the beliefs and disagree with the other (marked by the sharp peaks at the $(+1, -1)$ and $(-1, +1)$ points). The~parameters are: population size $N=100$, number of iterations $T$ = 50,000, and~connection type $K_0=-1/+1$, and~the noise value is $Z_A=0.01$.} \label{fig:EvolOfAtts6os}
\end{figure}

Assuming the most general setup, at~each time-step $t$, a~randomly chosen agent $i$ acquires the information, and~updates his/her attitudes according to Equation~(\ref{eq:att}). (For the flowchart of the algorithm, see Figure~\ref{fig:FlowChart}(a). The~source of information can be anything, such as public or social media, propaganda, government information, etc. As~it can be seen in Figure~\ref{fig:EvolOfAtts6os}a,b,d,e, proportionally to the level of exposure (iteration number $t$), the~attitudes tend to move towards the two extreme values, $+1$ and $-1$, either due to the experienced reassurance or due to the attempt to reduce cognitive dissonance (Equation~\ref{eq:att}). The~\emph{distribution} of the attitude values within the population evolves very similarly in case of the two attitudes, since both are governed by Equation~(\ref{eq:att}). (See Figure~\ref{fig:EvolOfAtts6os}a,b,d,e). At~high iteration numbers (indicating strong exposure to the news), around half of the population fully \emph{supports} Concept 1---marked by attitude values close to $+1$---while the other crowd---composed of those whose attitude values are close to $-1$---fully \emph{rejects} it. In~other words, the~population is \emph{polarized} with respect to Concept~1 (Figure~\ref{fig:EvolOfAtts6os}(a) and (d)). The~same applies to Concept~2 (Figure~\ref{fig:EvolOfAtts6os}(b) and (e)). In~case the type of connection ($K_0$) is negative (Figure~\ref{fig:EvolOfAtts6os} bottom row) %MDPI: the left bracket ``('' is missing, please check and revise. Answer: revised.
, the~two ``stable points''---adopted by the vast majority of the population---are $(+1, -1)$ and $(-1, +1)$, that is, where the two attitudes are reversed, either complete rejection of concept 1 and complete acceptance of concept 2 occurs, or~vice~versa. These are the two peaks in Figure~\ref{fig:EvolOfAtts6os}f. In~a symmetric manner, in~case the connection type, $K_0$ is positive, the~vast majority of the population will either completely support both concepts (one of the peaks will be at $(+1, +1)$), or~will completely reject both of them (the other peak will be at $(-1, -1)$), as~in Figure~\ref{fig:EvolOfAtts6os}c.
That is, independently of the type of connection, the~originally uniformly distributed attitudes will tend towards the extremities, meaning that the mere attempt to maintain a consistent belief system alone promotes the processing of attitudes tending towards extremities in case of being exposed to persistent information. Of~course, in~reality, it is not only one type of news that circulates within a community, but~many types, often with different messages and connotations, but~it is certainly an important---and so far overlooked---point, that this human drive (the urge to maintain consistent beliefs) alone has the capacity to push attitudes towards extremities---a phenomenon increasingly experienced in our increasingly connected~world. 

\begin{figure}[ht]
\centering
\includegraphics[width=\linewidth]{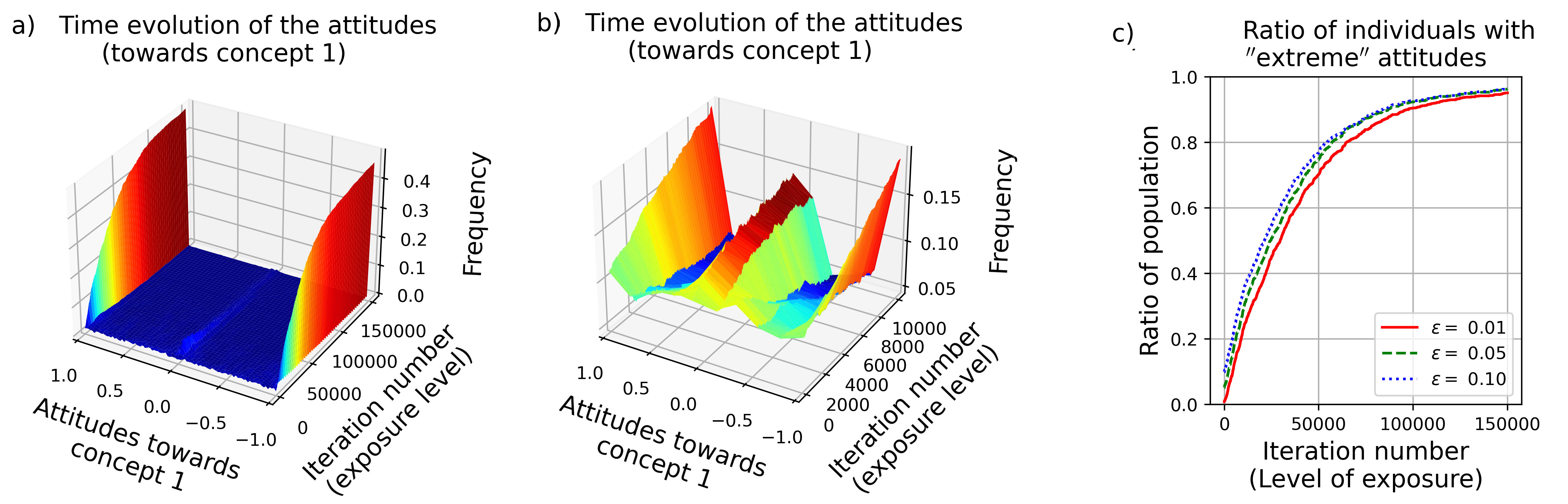}
\caption{A typical time evolution of an attitude value $a_{i,k}(t)$ as a function of time ($t$) for (\textbf{a}) large population ($N=1000$), and~(\textbf{b}) under limited exposure to news ($T$ = 10,000). In~this case, since in each time step 1 individual learns the news (out of the $N=1000$), on~average, each agent will have heard it 10 times at the end of the simulation. As~can be seen, for~such a level of exposure, developing a neutral standpoint (adopting attitude values close to zero) is a good ``strategy'' as well. However, this neutrality vanishes in case of more enduring circulation of the news. (\textbf{c}) Proportional to the level of exposure (iteration number $t$), the~ratio of the population holding ``extreme'' attitudes monotonically grows, independently of how ``extremity'' is defined (by the parameter $\epsilon$). 
The parameters are: population size $N=1000$, Number of iterations $T$ = 150,000 (except for sub-figure (\textbf{b}), on~which \mbox{$T$ = 10,000}), connection type between the concepts $K_0=-1$ and the noise value is $Z_A=0.01$.} \label{fig:EvolOfAtts}
\end{figure}

In Figure~\ref{fig:EvolOfAtts}c ``extremity'' is defined as ``being closer to +1 or $-1$ than a certain threshold value $\epsilon$''. Accordingly, if~$\epsilon = 0.01$, then attitudes between 0.99 and 1, and~attitudes between $-0.99$ and $-1$ will be considered as ``extreme''. Similarly, if~$\epsilon = 0.1$, then attitudes between 0.9 and 1, and~the ones between $-0.9$ and $-1$ will be considered as "extreme". Apparently, as~can be seen in Figure ~\ref{fig:EvolOfAtts}c, the~exact value of $\epsilon$ does not~matter.

Note the small peaks around near-zero values in Figure~\ref{fig:EvolOfAtts6os}a,b,d,e, at~small $t$ values. According to the simulations, \emph{in case of limited exposure to the news}, agents might also adopt neutral standpoints (marked by near-zero attitude values) in order to avoid cognitive dissonance. This phenomenon is highlighted in Figure~\ref{fig:EvolOfAtts}b. However, this is an unstable equilibrium point, since any further information regarding the given concept (appearing as noise $Z_A$ in Equation~(\ref{eq:att})) pushes the attitude value away from zero. (See also Supplementary Information, Figure S3).

\subsection{Finding Relief in New~Ideas}\label{sec:NewBelfResult}

As has already been mentioned, the~other ``basic strategy'' applied by people in order to reduce the unpleasant feeling of cognitive dissonance is to reinterpret the incoming information by placing it into a context in which the contradiction vanishes, or~even better, serves as a basis for reassurance~\cite{CDL:Festinger57}. For~example, a~doctor in his blog recollected memorable moments of the first year of the COVID-19 pandemic~\cite{DokiBlog}. He remembers that when he tried to convince his family members to take the vaccine, he received vehement rejection, which was settled by receiving the comment that ``You have good intentions, we know it. But~you do not see the reality, because~the ``big players'' leave you out from the party''. As it turned out, by~this they meant that the ``big players'' know perfectly well that the pandemic is a hoax, but~they use the everyday doctors---such as the one writing the blog---for their purposes, i.e.,~to force ``everyday people'' into take the unnecessary and harmful vaccine. In~this example, the~doctor is a positive concept in the eye of his relatives, but~the epidemic is negative (considered to be a hoax). When it turned out that the doctor considered the epidemic real (hence they should take the vaccine), he created a positive (supportive) relation between himself and the pandemic. This resulted in cognitive dissonance in the relatives, which was dissolved by adopting the new belief (about the ``party'' of the "big players"), which allowed the original attitudes to remain~unchanged.

\begin{figure}[ht]
\centering
\includegraphics[width=15cm]{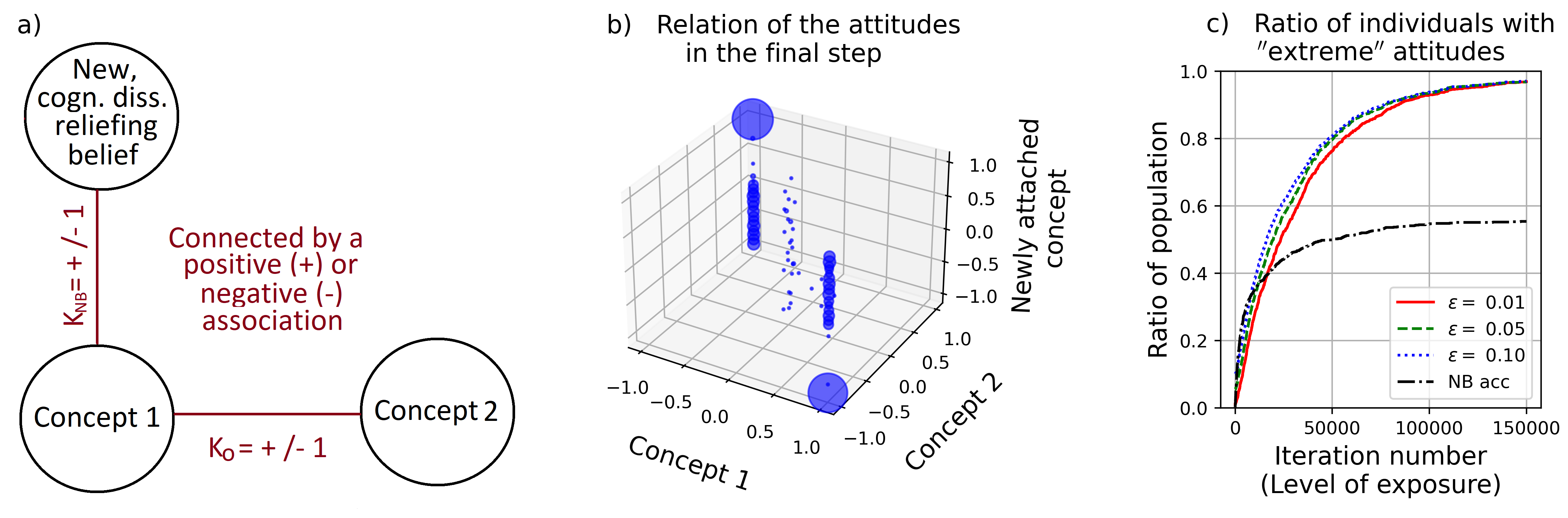}
\caption{Change of attitudes within a population due to the circulation of some news, connecting concepts 1 and 2. In~this case, agents \emph{might} adopt a new belief as well, in~case it reduces their cognitive dissonance. (\textbf{a}) The new belief can be connected to either or both concepts 1 and 2. (\textbf{b}) The ``stable configuration'' toward which the dynamics tends to (after 150,000 simulation steps). (\textbf{c}) The ratio of individuals holding ``extreme beliefs'', and~adopting the new belief (black semi-dotted line). ``Extreme attitudes'' are those closer to +1 or $-1$ than a certain threshold value $\epsilon$, such as 0.01, 0.05 and 0.1. As~it can be seen, the~ratio is largely independent of the exact value of $\epsilon$. The~parameters are: $N=1000$, $T$ = 150,000, $K_0=-1$, $K_{NB}=-1$ and $Z_A=0.01$. \label{fig:AccNewBel}}
\end{figure}

In the context of the present framework, this scenario can be represented by supplementing the original graph (including two nodes and an edge between them, as~in Figure~\ref{fig:BSNW}) with a new node, representing the new belief (see Figure~\ref{fig:AccNewBel}(a). The~new belief can be related to either of the original concepts, or~to both of them. As~an example, in~Figure~\ref{fig:AccNewBel}a, the~new belief is connected to Concept 1. The~type of connection, $K_{NB}$, can be either supportive or contradictory, similarly to the connection relating the two original concepts, $K_0$.

Figure~\ref{fig:AccNewBel}b depicts the ``stable configuration'' which the dynamics tends towards. As~has been shown already, in~the case of $K_0 = -1$, the~attitudes towards concept 1 and 2 tend to be antagonistic and extreme (marked by the attitude values accumulating in the $(-1, +1)$ and $(+1, -1)$ points on the $x-y$ plain), while in the case of $K_0 = +1$ (positive relation), the~attitudes towards concept 1 and 2 tend to be coincidental and also extreme (marked by the attitude values accumulating in the $(+1, +1)$ and $(-1, -1)$ points on the $x-y$ plain, see Supplementary Figure S5). The~vertical, $z$ axis depicts the attitude values towards the new, cognitive-dissonance-relieving belief; those who adopt it tend to develop an extreme relation towards this belief as well (in case of unceasing exposure). In~contrast, those for whom the approval of the new belief would create cognitive dissonance, simply reject its adoption. In~terms of the model, in~their case, the~edge $K_{NB}$ will simply not exist, and~hence the node representing this belief will not be connected to the belief network. The~two solid columns belong to these agents, depicting their original attitudes, which simply do not change throughout the simulation. (In case we stipulate that only those values are shown in the figure which participate in the belief system of an agent, these values could be omitted as well, but~for sake of clarity, in~Figure~\ref{fig:AccNewBel}b, we have kept them.) Furthermore, a~small vertical ``cloud'' can be seen in the middle of the chart, representing those who are neutral towards the original concepts, i.e.,~their attitude towards the new (cognitive-dissonance-relieving) belief can take any value.
Importantly, as~is apparent from Figure~\ref{fig:AccNewBel}, this mechanism pushes the attitudes towards extremities as~well. 

\section{Discussion}

The detailed methods by which humans perceive and make sense of the world---despite some eminent achievements~\cite{Searle10, Seth, Pinker, BelBrain}---is still to be understood. However, some basic characteristics have been elucidated by now, and~have become part of mainstream science as well~\cite{EBGoldsteinCognPsy, KCC, CogPsyTK}. One such characteristic is that in the human mind, beliefs are strongly interconnected, and~as such, no belief, concept or ``piece of information'' can exist on its own. Furthermore, in~case of new information, humans immediately attempt to interlock it in a coherent way, seeking for connections and support with already existing~beliefs. 

Accordingly, the~novelty of the present model lies not in ``assuming'' the above-mentioned two human characteristics---since they are well-studied and widely accepted by main-stream science~\cite{CsanyiHied, CDL:Festinger57, CognDissA}---rather, it lies in their mathematical formulation and incorporation into agent-based~models.

There are two more further points worthy of consideration related to the model:

\begin{enumerate}
\item[(i)] Real belief systems have a tremendous amount of elements (instead of two or three), that are interconnected and embedded into each other in a complicated manner~\cite{CsanyiHied, BelBrain}, and, accordingly, the~``optimization process''---the attempt to minimize the contradictions among the components---refers to the \emph{entire} system. From~ a physicist's point of view, this process is in close relation to physical structures aiming to reach an energy minimum. 
In this approach, ``different realities''~\cite{Scout} can be different local energy minimums of similar systems.
However, it is imperative to understand the \emph{elementary relation} between \emph{two} elements of the system before considering the entire structure. The~present manuscript focuses on this elementary relation. 
Graph representation is important because, and~only because, it serves as a mathematical tool for handling interrelated entities (which are the ``beliefs'' or ``concepts'' in our case). Since in the human mind a vast amount of concepts and beliefs are interrelated densely and intricately, any of its graph representations must also assume a vast amount of intricately interrelated (linked) nodes. However, from~the viewpoint of the present study, the~specific type of the graph does not play any role, because~we focus on the elementary process altering the characteristics of two nodes (namely the ``attitude values'') due to a newly appearing link between them. (If a link appears, it is due to a certain piece of information connecting the two, originally unconnected beliefs/concepts). The~nodes whose values alter are selected by the link (representing a piece of information).

\item[(ii)] The present model does not assume that the repeated information is \emph{exactly} the same, only that the \emph{type of connection} between two concepts (say a political party and a public issue, such as immigration or environmental topics) is tenaciously either positive or negative. Hence, it also explains how attitudes can become extreme due to the continuous repetition of information, and~as such, it serves as a complementary explanation~\cite{CentolaB1} for the reason why, throughout history, the~most diverse regimes found it useful to repeat the same messages over and over again (despite the fact that everybody had already heard them many times).
\end{enumerate}

Furthermore, the~present model has some additional results as well, which are yet to be studied. Specifically, according to the results, the~attitudes a certain type of news or information triggers depend on the \emph{intensity} of the exposure. More precisely, in~case of limited exposure, people tend to develop a centralist attitude first (which is an unstable equilibrium point), which, in~case of persistent news-circulation, give way to extreme stances. The~dynamics under \emph{limited} exposure to news was not studied extensively in the present~manuscript.

The ambition of the paper was to call attention to certain human traits that have not yet been incorporated into current computational models aiming to simulate opinion dynamics in human communities. From~this perspective, the~main point is the naturalness by which polarization can emerge, despite the fact that the model incorporates only minimal assumptions which are considered to be part of well-established, main-stream scientific~results.

\section{{Appendix: Flowchart and Parameters}}

\begin{figure}[ht]

\centering
\includegraphics[width=15cm]{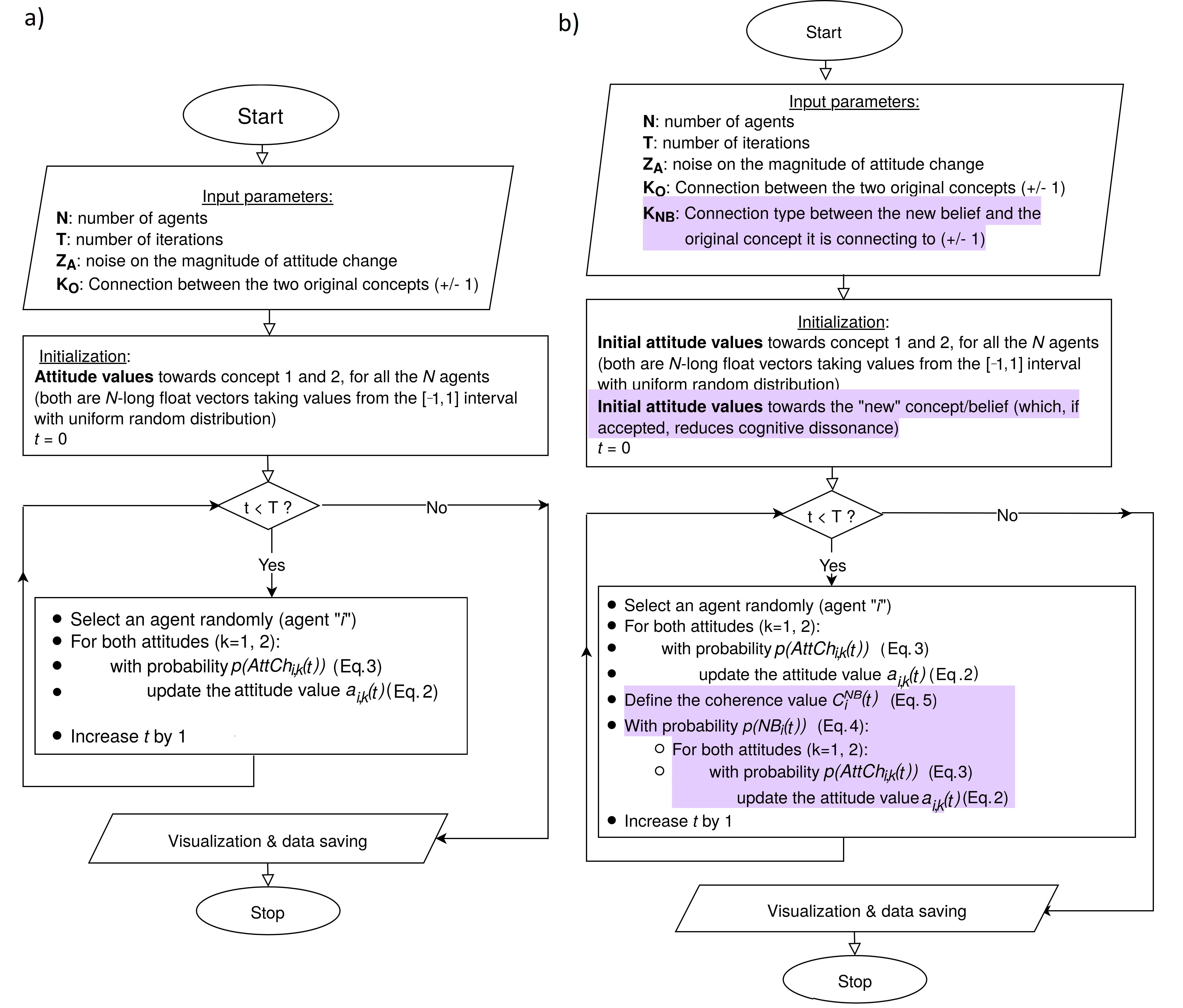}
\caption{Flowchart of the algorithms. {(\textbf{a})} is the algorithm of the case discussed in the Section~\ref{sec:ReevalBSResults}. {(\textbf{b})} depicts the algorithm detailed in the Section~\ref{sec:NewBelfResult}. The~main difference -- highlighted with purple -- is that in the latter case, agents might adopt a new belief as well, in~case it decreases their cognitive~dissonance.\label{fig:FlowChart}}
\end{figure}

The simulation was written in Python. Figure~\ref{fig:FlowChart} shows the flowchart of the algorithm which is enough to replicate the results. However, the~source code of the simulation can be found on CoMSES, a~public computational model library as well~\cite{SourceCode}. Figure~\ref{fig:FlowChart}a shows the algorithm for the results explained in Section~\ref{sec:ReevalBSResults}, and~Figure~\ref{fig:FlowChart}b depicts the algorithm for the case when people might accept a new, cognitive-dissonance-relieving belief as well, the~case explained in Section~\ref{sec:NewBelfResult}.

Since this is a minimalist model, altogether 5 parameters were used, which are over-viewed in Table~\ref{tab:params}. The~robustness of the results have been validated for a wide range of parameters, for~which the results can be found in the Supplementary~Material.

\begin{table}[ht!]
\centering
\begin{tabular}{|c|c|c|}
\hline
\textbf{Notation}	& \textbf{Meaning}	& \textbf{Values} \\
\hline
N	& Number of individuals (population size)	& 1000 (SI: 100)\\
\hline
T	& Number of iterations (level of exposure) & 100,000--300,000\\
\hline
$Z_A$ & Noise on the magnitude of attitude change & $\in [-0.01, +0.01]$ \\
 &  & SI: $\in [-0.05, +0.05]$, \\
 &  & $\in [-0.05, +0.05]$ \\
 &  & $\in [-0.1, +0.1]$ \\
 &  & $\in [-0.2, +0.2]$ \\
 &  & $\in [-0.5, +0.5]$ \\
\hline
$K_O$ & Positive or negative: Type of the connection between the two original concepts & $+1$ or $-1$ \\
\hline
$K_{NB}$ & Type of the connection between the newly accepted concept and the one it is connected to. & $+1$ or $-1$ \\
\hline
\end{tabular}
\caption{Summary of the model parameters. Left column: Nomination used in the manuscript. Middle column: description, and~right column: values used in the~simulation.\label{tab:params}}
\end{table}

%Bibliography
\bibliographystyle{unsrt}  
\bibliography{refek}

\section*{Acknowledgements}

The research was partially supported by the Hungarian National Research, Development and Innovation Office (grant no. K 128780).

\section*{Author contributions statement}

A. Z. has designed the model, developed the code, analysed the results and wrote the manuscript. 

\section*{Data availability}

The flow-chart of the model is presented in Figure~\ref{fig:FlowChart} (in the "\emph{Methods}" Section). The source code of the simulation is available at the \emph{CoMSES} public computational model library (\cite{SourceCode}).

\section*{Competing interests}

The author declares no competing interests.

%%%%%%%%%%%%%%%%%%%%%%%%%%%%%%%%%%%%%%%%%%%%%%%%%%%%%%%%%%%%%%%%%%%%%%%5
%%%%%%%%%% Merge with supplemental materials %%%%%%%%%%
% \pagebreak
\vspace{2cm}
\begin{center}
\textbf{\large Supplementary Information: Opinion polarization in human communities can emerge as a natural consequence of beliefs being interrelated}
\end{center}
%%%%%%%%%% Merge with supplemental materials %%%%%%%%%%
%%%%%%%%%% Prefix a "S" to all equations, figures, tables and reset the counter %%%%%%%%%%
\setcounter{equation}{0}
\setcounter{figure}{0}
\setcounter{table}{0}
\setcounter{page}{1}
\makeatletter
\renewcommand{\theequation}{S\arabic{equation}}
\renewcommand{\thefigure}{S\arabic{figure}}
% \renewcommand{\bibnumfmt}[1]{[S#1]}
% \renewcommand{\citenumfont}[1]{S#1}
%%%%%%%%%% Prefix a "S" to all equations, figures, tables and reset the counter %%%%%%%%%%

The aim of the present Supplementary Information is to study the robustness and statistical behaviour of the results presented in the main text. The first subsection provides an analysis of the statistical behaviour of the simulation. In the second part, the effects of the parameters are surveyed. And finally, in the last subsection, entitled "Involving the hierarchical nature of beliefs", we study the behaviour of the model under the assumption that more deeply embedded (that is: \emph{more central}, or \emph{higher-ranking}) beliefs are more difficult to change.

\subsection*{Statistical behaviour}

\begin{figure}[ht!]
\centering
\includegraphics[width=\linewidth]{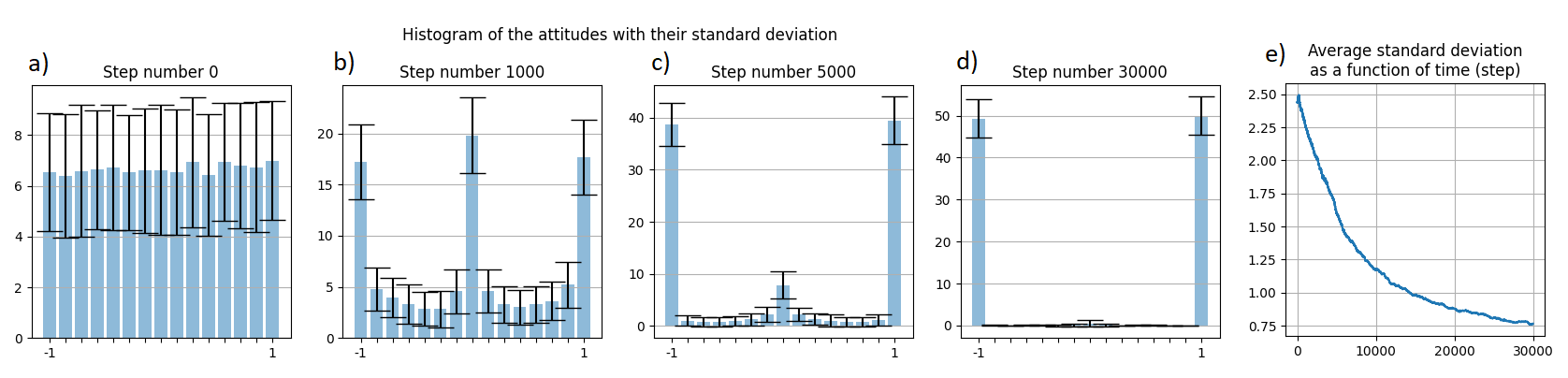}
\caption{Histogram of the attitude values, along with their standard deviations, for four time-steps: $t=0$ (sub-figure \textit{\textbf{a}}), $t=1,000$ (\textit{\textbf{b}}), $t=5,000$ (\textit{\textbf{c}}), and $t=30,000$ (\textit{\textbf{d}}). Sub-figure \textit{\textbf{e}} depicts the average of the standard deviations along the entire run, from $t=0$ to $t=30,000$. As it can be seen, with the growth of $t$, the standard deviations of the histograms values decrease, leading to the conclusion that the long-term outcome of the simulation -- basically: its predictions -- are independent of the initial values. These histograms can be considered as slices of the 3D histograms presented in various places throughout the manuscript (for example in sub-figures \textit{\textbf{a}} and \textit{\textbf{b}} in Figure 2 in the main text), and the Supplementary Information. 
The parameters are: number of runs: 200. Number of bins: 15. Number of iterations: 30,000. Initial distribution: uniform, taking values from the [-1, 1] interval. Number of agents, $N=100$. Time-steps: $t=0$ (sub-figure \textit{\textbf{a}}), $t=1,000$ (sub-figure \textit{\textbf{b}}), $t=5,000$ (sub-figure \textit{\textbf{c}}) and $t=30,000$ (sub-figure \textit{\textbf{d}}).}
\label{fig:SIStatAnal}
\end{figure}

In the initial step of the simulation, the attitude values are set according to a certain distribution. In the results presented in the main text, the initial distribution is uniform, taking values from the $[-1, 1]$ interval. This setting calls for two questions: (i) How does the dynamics alter in case the initial distribution is not uniform?, and (ii) How does the stochastic nature (originating from the random numbers) affect the simulation results? In order to answer the first question, we have run the simulation with various types of initial distribution. The results are summarized in Supplementary Figure~\ref{fig:SIDistrTabla} in the subsection entitled "The initial distribution of the attitude values". 
In order to answer the second question, we have executed the simulation 200 times and calculated the histogram of the attitude values, along with their standard deviation, at each time-step $t$. Sub-figures \textit{\textbf{a}} to \textit{\textbf{d}} in Supplementary Figure~\ref{fig:SIStatAnal} shows the results for four time-steps: the initial one: $t=0$, and for $t=1,000$, $5,000$ and $30,000$. As it can be seen, with the growth of $t$, the standard deviations of the histograms values decrease, leading to the conclusion that with the progress of the simulation, "as time goes by", the effect of the initial randomness decreases. In other words, the long-term outcome of the simulation is well-defined and independent of the actual values in the initial step. Sub-figure \textit{\textbf{e}} depicts the average of the standard deviations for each time-step $t$, leading to the same conclusion. These results were calculated for the attitude values towards concept 1, but since the dynamics of the attitude values are governed by the same rules for both concepts (defined by Equation 2 in the main text), these results are valid for concept 2 as well.
\vspace{1cm}

\subsection*{Effect of the parameters}

The model -- as designed to be a minimal model -- contains only a couple of parameters, summarized in Table 1 in the main text (repeated hereinafter). In the followings, the effect of their actual values will be inspected.

\begin{table}[ht!]
\centering
\begin{tabular}{|c|c|c|}
\hline
\textbf{Notation} & \textbf{Meaning} & \textbf{Values} \\
\hline
N & Number of individuals (population size) & 100, 1000 \\
\hline
T & Number of iterations (level of exposure) & 100,000-300,000 \\
\hline
$Z_A$ & Noise on the magnitude of attitude change & 0.01, (SI: 0.01-0.5) \\
\hline
$K_O$ and $K_{NB}$ & Type of the connection between concepts (supportive or conflicting) & $+1$ or $-1$ \\
\hline
\end{tabular}
\caption{\label{tab:paramsSI}Summary of the model parameters. Left column: Nomination used in the manuscript. Middle column: description, and right column: values used in the simulation. See also Table 1 in the main text.}
\end{table}

\subsubsection*{Number of agents, N}

\begin{figure}[ht!]
\centering
\includegraphics[width=\linewidth]{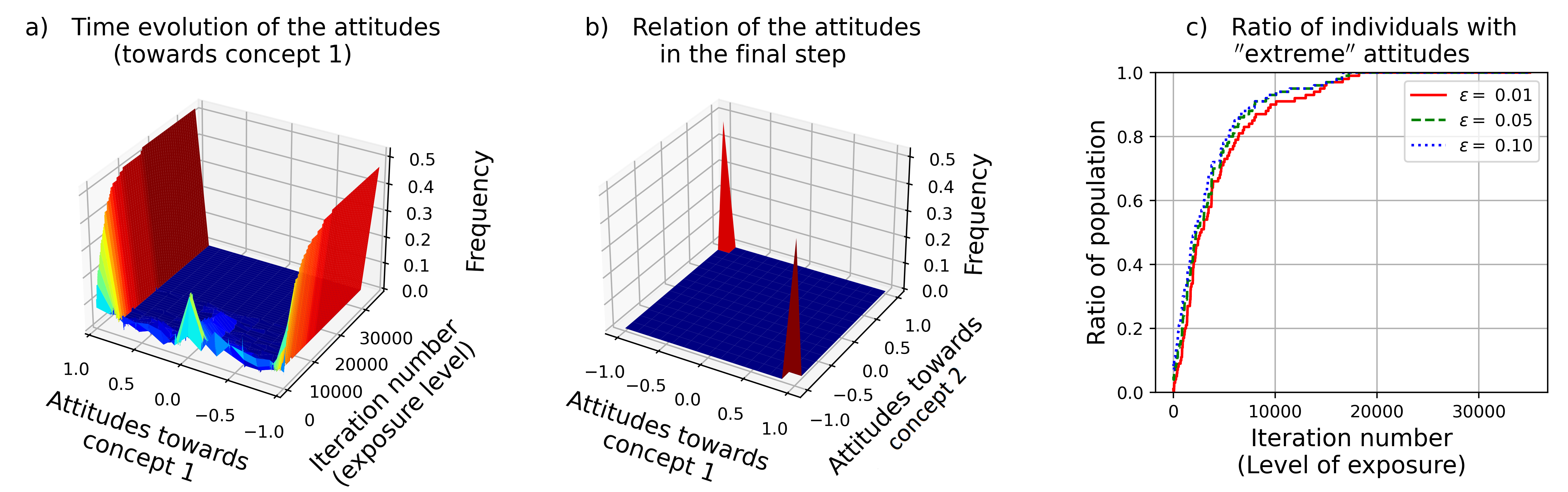}
\caption{Simulation results for $N=100$ agents, that is, for one scale smaller population than the one studied in the main text. \textit{\textbf{a)}} The evolution of attitudes within a population, due to exposure to a news: similarly to the results reported in the main text, as the level of exposure intensifies -- measured by the iteration number $t$ --  the attitude values tend towards extremities, marked by $+1$ and $-1$. At the beginning of the process (at smaller iteration numbers, representing lower exposure) maintaining zero-near attitude values is also a good strategy for avoiding cognitive dissonance. \textit{\textbf{b)}} At the end of the simulation, the vast majority of the population adopts extreme attitudes, marked by the sharp peaks at $(+1, -1)$ and $(-1, +1)$.
\textit{\textbf{c)}} Proportional to the level of exposure (iteration number $t$), the ratio of population holding extreme attitudes monotonically grows. 
Parameters: $N=100$, $Z_A=0.01$, $T = 35,000$, News connection type: negative ($K_0 = -1$), and finally, a parameter effecting only the visualization: number of bins = 15 in sub-figures \textit{\textbf{a}} and \textit{\textbf{b}} (instead of the 100, applied for figures in the main text).}
\label{fig:SIMeret100}
\end{figure}

Equations (1) to (5) (in the main text) govern the dynamics of the attitude values. Since all of these equations include data describing only the focal individual (agent $i$), individuals develop their attitudes basically independently from each other. This originates from the fact that in the present model, the source of the information does not matter. Accordingly, the exact number of the population ($N$) has only a statistical effect (as higher number of agents ensures more precise statistics). The results provided in the main text are generated assuming $N=1000$, but, as Supplementary Figure~\ref{fig:SIMeret100} demonstrates, the main features of the dynamics are perfectly visible on smaller populations as well (in which $N=100$). (The only way by which the population size matters is the speed of the dynamics, since in smaller populations each agent gets to be selected more frequently -- but this feature can be counteracted by longer runs (higher iteration numbers )).

\subsubsection*{Number of iterations (level of exposure), T}

As briefly mentioned beforehand, at low level of exposure (expressed by small values of the $t$ iteration number), maintaining neutral attitudes towards one or both concepts also results in the avoidance of cognitive dissonance (See Supplementary Figure~\ref{fig:SIRovidN100}). This feature -- observed in real-life cases as well -- manifests itself in the frequent occurrence of zero-near values (see sub-figure \textit{\textbf{a}}). However, prolonged exposure to the news/information destroys this neutrality, and pushes the agents towards extremity. The reason behind this phenomenon is that the attitude values immediately start to tend towards the extreme values as the agent tilts towards either direction -- in other words, it is an unstable equilibrium point. Accordingly, the higher the noise, the less stable this attitude is (see also Supplementary Figure~\ref{fig:SIZaj})

\begin{figure}[ht!]
\centering
\includegraphics[width=\linewidth]{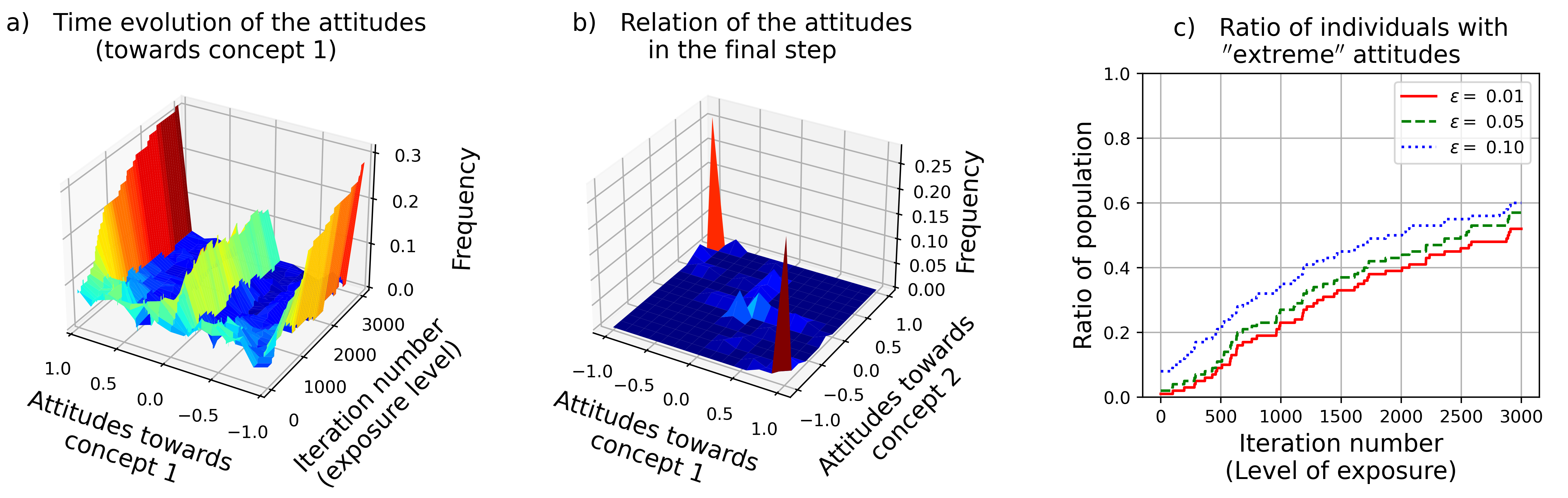}
\caption{The evolution of attitudes within a population, due to \emph{limited} exposure to news, achieved by limiting the number of iterations in $T = 3000$. \textit{\textbf{a)}} In case of limited exposure to news, zero-near values are also common, that is, "centralist" attitudes also appear in high number, since they ensure the avoidance of cognitive dissonance as well. However, since due to the smallest deviation from $0$, the attitudes start to tend towards extremities, this is an unstable equilibrium point, and, accordingly, in case of persistent exposure to news and/or experiencing higher levels of noise values, it vanishes, giving rise to extreme stances.} \textit{\textbf{b)}} While a considerable ratio of the population adopts extreme attitudes (marked by the peaks at $(+1, -1)$ and $(-1, +1)$), individuals also tend to take a neutral stance (characterized by zero-near values) for both concepts. \textit{\textbf{c)}} As the exposure to the news intensifies, the ratio of population holding extreme attitudes monotonically grows. Parameters: $N=100$, $Z_A=0.01$, $T = 3,000$, News connection type: negative ($K_0 = -1$), and finally, a parameter effecting only the visualization: number of bins = 15 in sub-figures \textit{\textbf{a}} and \textit{\textbf{b}}.
\label{fig:SIRovidN100}
\end{figure}

\subsubsection*{Noise, $Z_A$}

\begin{figure}[ht!]
\centering
\includegraphics[width=.8\linewidth]{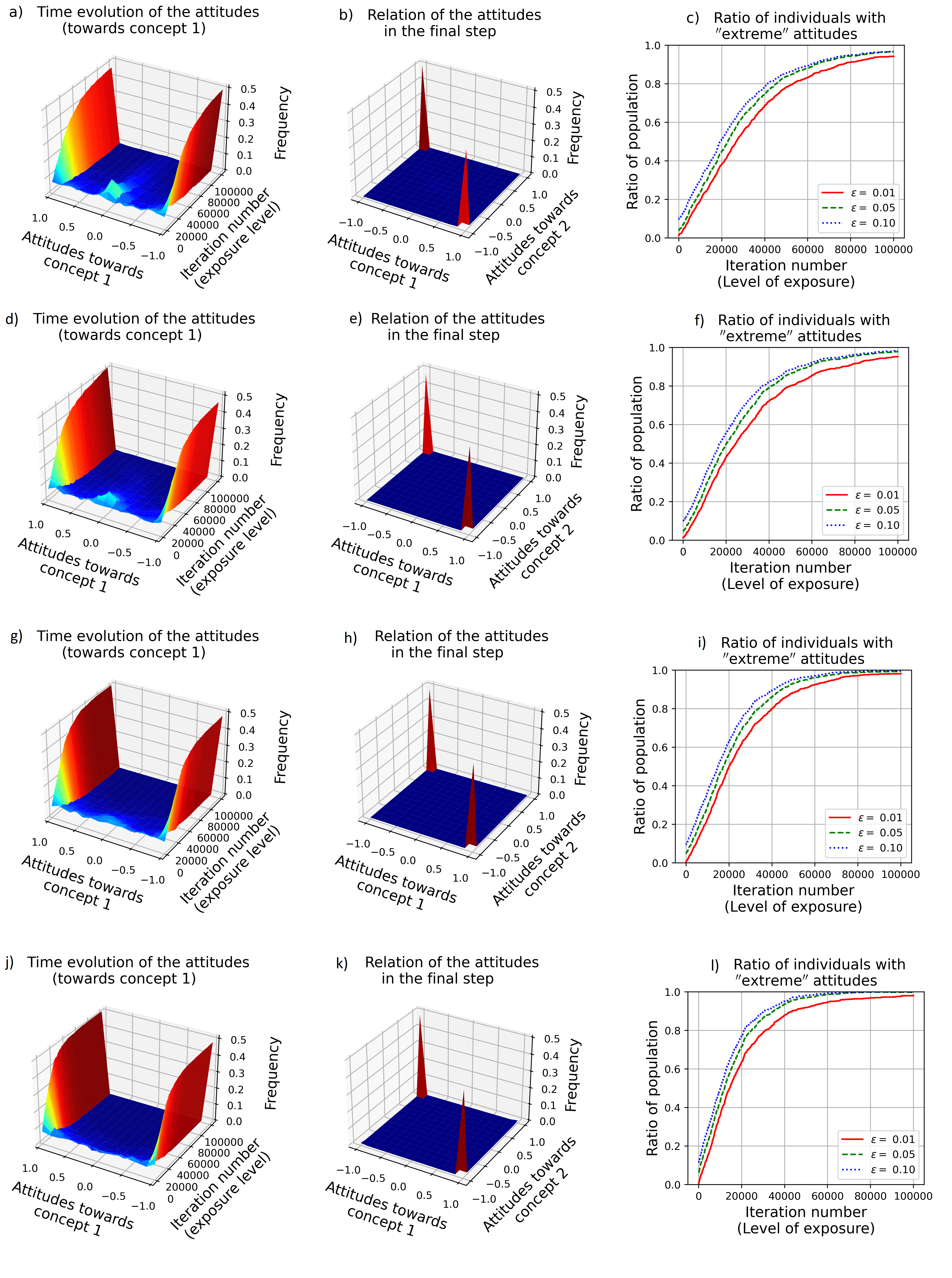}
\caption{Simulation results with four different noise values: $Z_A = 0.05$ (on sub-figures \textit{\textbf{a}} to \textit{\textbf{c}}, top row), $Z_A = 0.1$ (on sub-figures \textit{\textbf{d}} to \textit{\textbf{f}}, second row), $Z_A = 0.2$ (on sub-figures \textit{\textbf{g}} to \textit{\textbf{i}}, third row) and $Z_A = 0.5$ (on sub-figures \textit{\textbf{j}} to \textit{\textbf{l}}, bottom row.) First column: The time evolution of the attitude-histograms; Second column: attitude values in the final step, and third row: ratio of individuals with extreme attitude values. As it can be seen, noise does not effect the long-term outcome of the simulation results, as the attitude values clearly tend towards the (-1, 1) and (1, -1) stable points. However, noise does have an important effect on the short run: the higher the noise is, the less stable the neutral (zero-near) standpoint are, and, accordingly, they vanish more quickly. The parameters are: $N=1000$, $T = 100,000$, news connection type: negative ($K_0 = -1$), and the number of bins = 15}.
\label{fig:SIZaj}
\end{figure}

Noise is a central element in all social and biological models. In the present approach, two types of noises are included: $\rho$ and $Z_A$. The fist one, $\rho$ has effect only in case the hierarchical nature of beliefs are also considered (see section "Involving the hierarchical nature of beliefs"). The second one, $Z_A$ defines the maximal value of attitude change (see also Eq. (2) in the main text). The bigger the $Z_A$, the larger is the \emph{magnitude} of the random change of the updated attitude values. This random variation can be either positive or negative, allowing attitudes to change signal. Supplementary Figure~\ref{fig:SIZaj} shows the simulation results assuming four different noise levels: $Z_A = 0.05$ (top row), $Z_A = 0.1$ (second row), $Z_A = 0.2$ (third row) and finally, $Z_A = 0.5$ (bottom row). These are to be compared to $Z_A = 0.01$, assumed in the simulations presented in the main text. The most important effect of the noise is the elimination of the neutral stand-points: despite the fact that zero-near attitudes protect the agent from experiencing cognitive dissonance, the higher the noise is, the less stable this equilibrium point is, since noise tilts the attitude values away from 0.

\subsubsection*{Type of connection, $K_O$ and $K_{NB}$}

\begin{figure}[ht!]
\centering
\includegraphics[width=\linewidth]{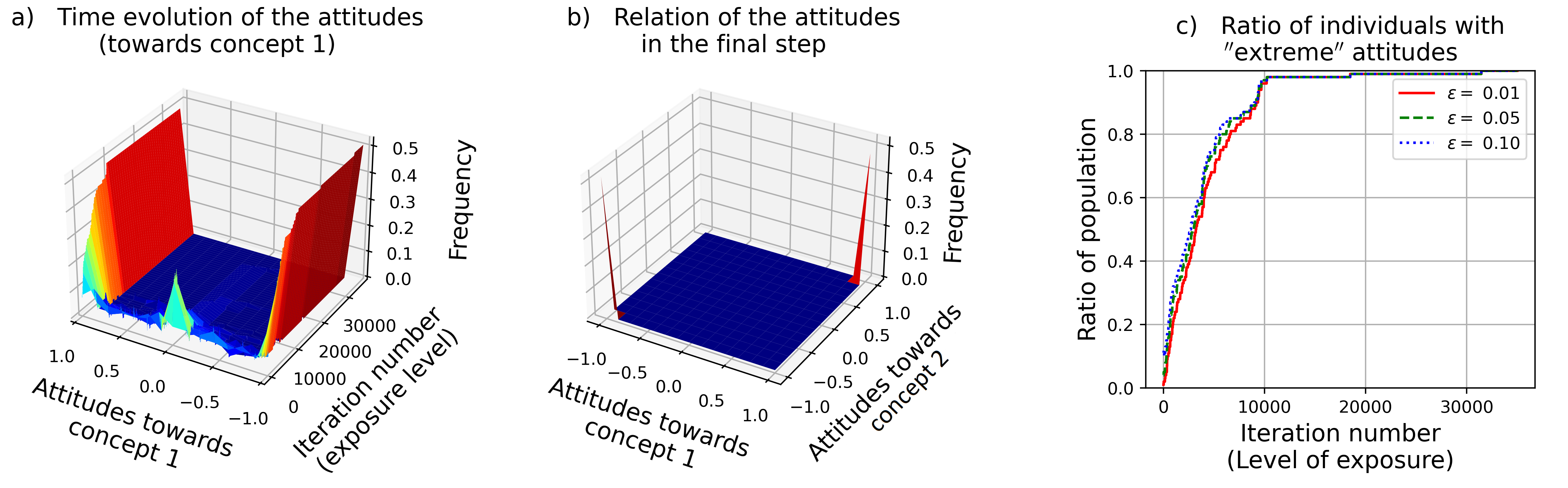}
\caption{
The effect of the choice of $K_0$, the type of connection. \textit{\textbf{a)}} The evolution of attitudes within a population, due to exposure to a news: similarly to the results reported in the main text, as the level of exposure intensifies -- measured by the iteration number $t$ --  the attitude values tend towards extremities, marked by $+1$ and $-1$.
\textit{\textbf{b)}} At the end of the simulation, the vast majority of the population adopts extreme attitudes, \emph{but in this case, they coincide, instead of opposing each other}, marked by the sharp peaks at $(+1, +1)$ and $(-1, -1)$.
\textit{\textbf{c)}} Proportional to the level of exposure (iteration number $t$), the ratio of population holding extreme attitudes monotonically grows. 
Parameters: $N=100$, $Z_A=0.01$, $T = 35,000$, News connection type: positive ($K_0 = +1$), and finally, the number of bins = 15 in sub-figures \textit{\textbf{a}} and \textit{\textbf{b}}.}
\label{fig:SIPozConnTypeN100}
\end{figure}

In case the way the news connecting the concepts is not negative but positive ($K_0 = +1$ instead of $-1$), then -- according to the expectations -- attitudes towards the connected concepts coincide with each other, instead of opposing one an other (see sub-figure~\textit{\textbf{b}} in Fig.~S\ref{fig:SIPozConnTypeN100}).

This follows from the fact that the level of coherence individual $i$ experiences at time-step $t$, $C_i^t$, depends only on the original attitudes, $a(1)_i^t$, $a(2)_i^t$ and on the connection type $K_0$, as $C_i^t = a(1)_i^t \cdot a(2)_i^t \cdot K_0$ (See also Eq. 1 in the main text). As it can be seen in Figure~S\ref{fig:SIPozConnTypeN100}, apart from this effect, the exact choice of $K_0$ (whether it is +1 or -1) does not have any other effect on the simulation results.

\subsubsection*{The initial distribution of the attitude values}

For the results presented in the main text, the initial attitude values (that is: the way agents relate to one or the other concept) are set randomly, according to a uniform distribution, taking values from the $[-1, 1]$ interval. The effect of this stochasticity, originating from the randomly determined initial values, has been analysed in the first part of the Supplementary Information, in the section entitled "Statistical behaviour". In the present section simulation results are shown for different settings of initial attitude values. More specifically, four cases are being compared (all taking values from the $[-1, 1]$ interval): (i) uniform distribution, same as in the main text, see Supplementary Figure \ref{fig:SIDistrTabla}~\textit{\textbf{a}} (ii) Gauss distribution with $\sigma = 0.4$, (see Fig.~S\ref{fig:SIDistrTabla}~\textit{\textbf{e}}), (iii) Gauss distribution with $\sigma = 0.2$, (see Fig.~S\ref{fig:SIDistrTabla}~\textit{\textbf{i}}), and (iv) constant zero (Fig.~S\ref{fig:SIDistrTabla}~\textit{\textbf{m}}). The central question is that how these initial differences affect the dynamics of the simulation and its long-term outcome? As it can be seen on the third column of Fig. S\ref{fig:SIDistrTabla} (sub-figures ~\textit{\textbf{c}}, ~\textit{\textbf{g}}, ~\textit{\textbf{k}} and ~\textit{\textbf{o}}), in all cases, the attitudes tend toward extremities, marked by the sharp peaks at the $[-1, 1]$ and $[1, -1]$ points. At the same time, the dynamics itself is different, as the more close the initial values are to zero, the longer it takes for these neutral standpoints to evolve extreme (see the 2$^{nd}$ column in Fig.~S\ref{fig:SIDistrTabla}). Note, that in case all initial attribute values were set to be zero (bottom row), the simulation run for 300,000 steps, in contrast with the 150,000, set for the first three cases. At the same time, this process speeds up by applying higher nose levels (see the section entitled "Noise, $Z_A$", and Supplementary Figure~\ref{fig:SIZaj})

\begin{figure}[ht!]
\centering
\includegraphics[width=\linewidth]{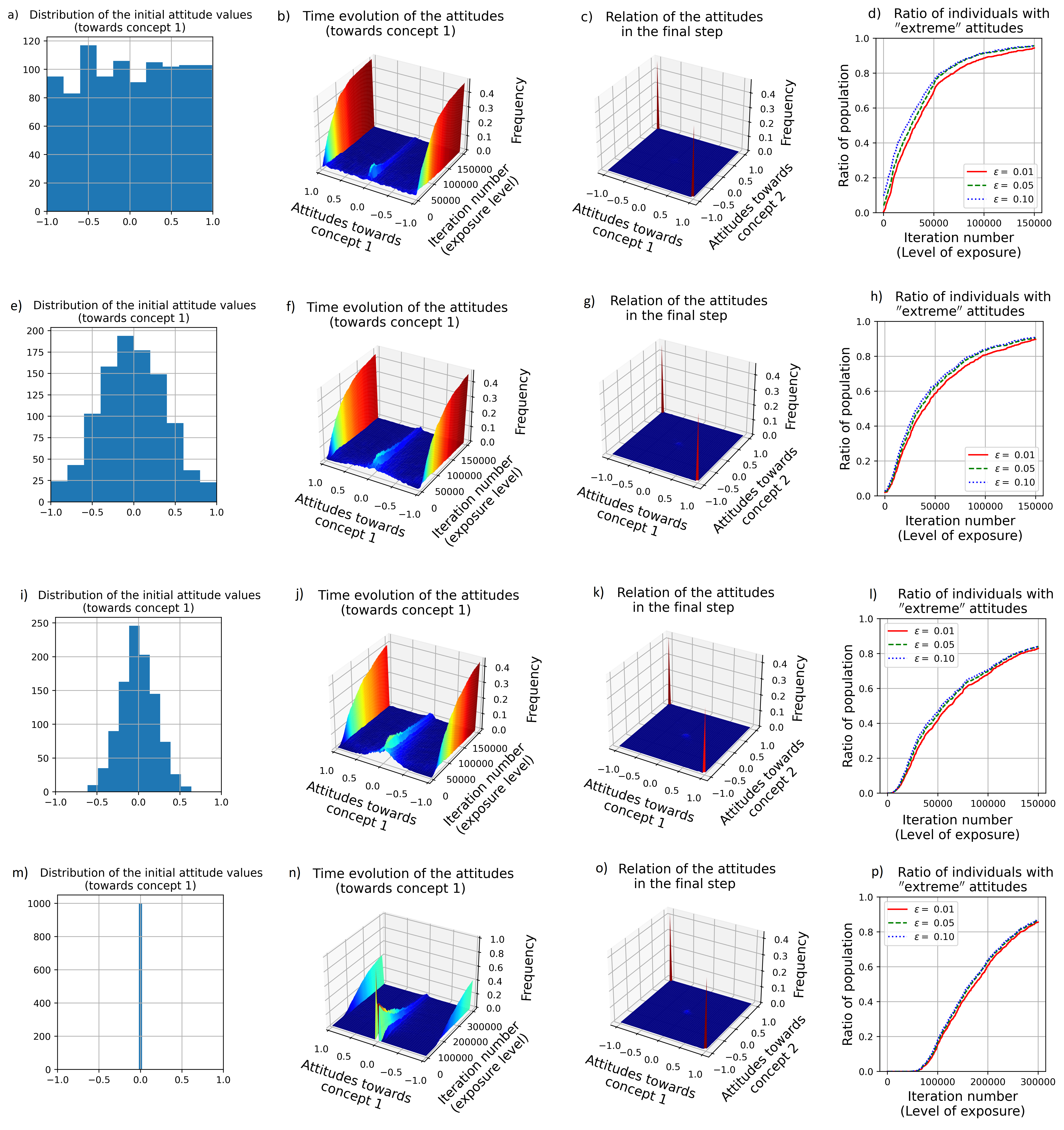}
\caption{Simulation result with four different initial attitude value distributions: top row: uniform distribution (sub-figure \textit{\textbf{a}}), 2$^{nd}$ row: Gauss distribution with $\sigma=0.4$ (sub-figure \textit{\textbf{e}}), 3$^{rd}$ row: Gauss distribution with $\sigma=0.2$ (sub-figure \textit{\textbf{i}}), and bottom row: constant zero distribution (sub-figure \textit{\textbf{m}}). All simulations tend towards the $[-1, 1]$ and $[1, -1]$ stable points on the long run (3$^{rd}$ column), but the dynamics are different (2$^{nd}$ column), as the more close the initial attitude values are to zero, the longer it takes for these neutral standpoints to evolve into extremities. The parameters are: $N=1000$, $Z_A=0.01$, $T = 150,000$ ($300,000$), News connection type: $K_0 = -1$, and finally, the number of bins = 50.}
\label{fig:SIDistrTabla}
\end{figure}

\vspace{.9cm}

\subsection*{Involving the hierarchical nature of beliefs}

As mentioned in the main text, human beliefs are organized in a hierarchical manner, meaning that there are more and less central convictions, having bigger or smaller effect on other beliefs. Typically, those that are more close to the \emph{identity} are higher in rank, and, accordingly, more difficult to change. 

The robustness of the model results with respect to this consideration has been checked by assuming that more embedded attitudes -- that is, those that are more close to +1 or -1 -- are more difficult to change. More precisely, the \emph{maximal magnitude} of the change has been set in a way that the more close an attitude is to abs(1), the smaller is the maximal change it can undergo. 

Recalling the equation from the main text (Eq. 2), defining the new attitude $a_i^{t+1}$ of agent $i$ at time step $t+1$, due to experiencing a cognitive dissonance (or reassurance) $C_i^t$, with noise level $Z_A$:

\begin{equation}\label{eq:attSI}
a(k)_i^{t+1} = sign(a(k)_i^t) \cdot ( \ \mid a(k)_i^t \mid + \rho \cdot C_i^t ) + Z_A\
\end{equation}

We see that the attitude value changes proportionally to the experienced "coherence" $C_i^t$ (which is usually referred to as "cognitive dissonance" in case it is negative, and "reassurance", in case it is positive). In the results presented in the main text, $\rho$ is simply a random number taken from the $[0, 1]$ interval with uniform distribution, that is, 

$$ \rho = R = Rand(0,1) $$

In order to incorporate the hierarchical property of beliefs into the model, $\rho$ has been expanded to have two components: (1) the above random value (taken from the $[0, 1]$ interval with uniform distribution), and (2) an "amplitude" value $M$ with which the above random value is multiplied with: 

$$ \rho = R \cdot M $$

This maximal value $M$ is defined by a simple Gauss function, and adjusted by its standard deviation $c$ in the following way: 

$$ M = e^{\frac{-a_i^t}{c^2}} $$

\begin{figure}[ht!]
\centering
\includegraphics[width=\linewidth]{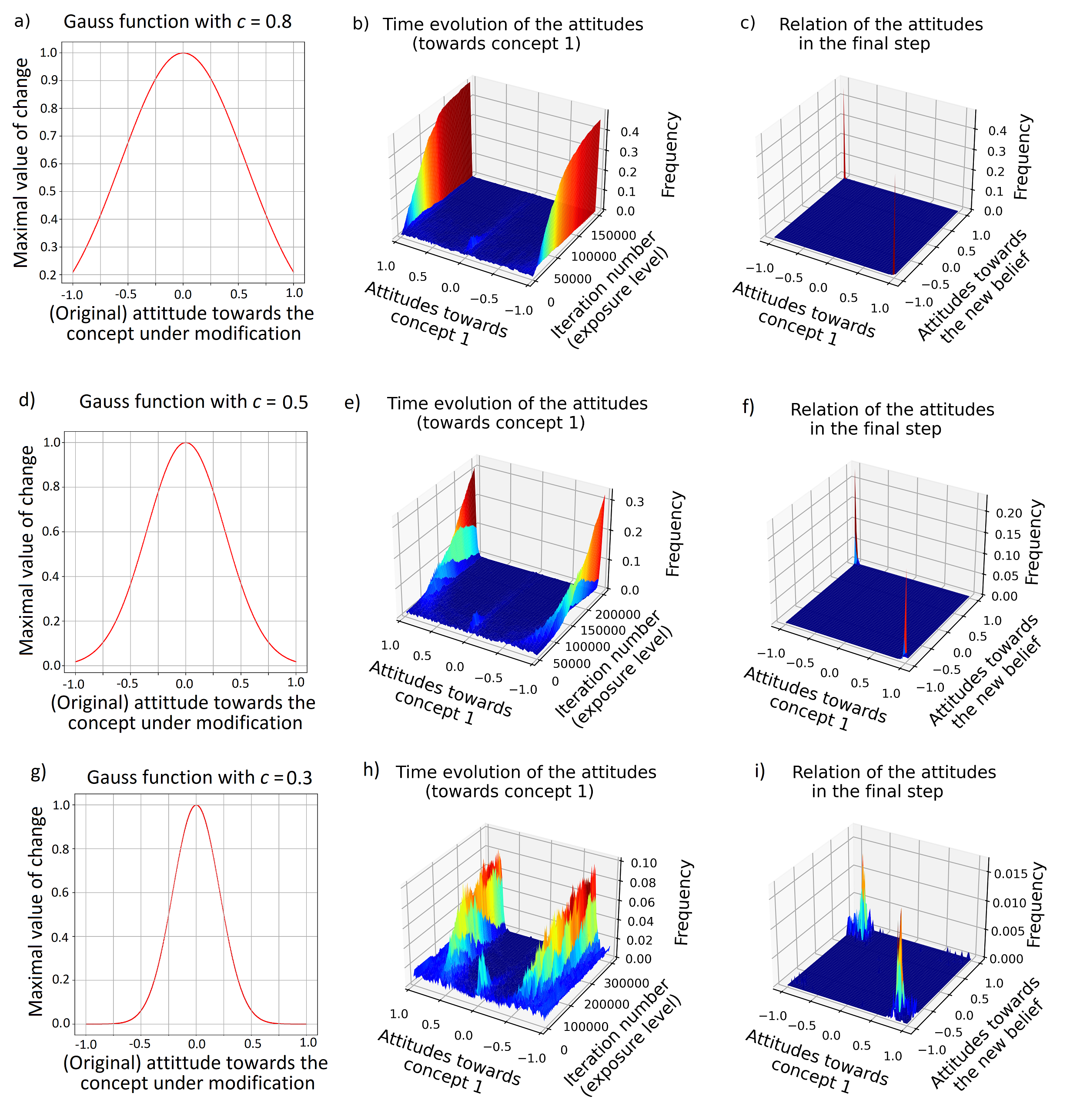}
\caption{Incorporating the hierarchical nature of beliefs. The change of attitudes within a population due to circulation of news for 3 different values of $c$: $c=0.8$ (top row), $c=0.5$ (middle row), and $c=0.3$ (bottom row). The smaller the $c$, the more pronounced is the hierarchical feature of the belief system (that is, the more difficult it is to modify high-ranking beliefs). \textit{\textbf{a)}}, \textit{\textbf{d)}} and \textit{\textbf{g)}}: the maximal value of attitude change $M$ as a function of $c$. \textit{\textbf{b)}}, \textit{\textbf{e)}} and \textit{\textbf{h)}}: The evolution of one of the attitudes, due to prolonged exposure to the news. The main claim remains unchanged: as the exposure to the news intensifies (marked by the iteration number $t$), attitude values drift towards the extremities. Furthermore -- as a new result -- the smaller is the $c$ is (that is, the more pronounced the hierarchical effect), the more are the "centralist" (near-zero) attitudes in case of moderate exposure (small $t$ values). \textit{\textbf{c)}}, \textit{\textbf{f)}} and \textit{\textbf{i)}}: The vast majority of the population adopts extreme attitudes, marked by the peaks at $(-1, +1)$ and $(+1, -1)$, in case of enduring exposure to the news.}
\label{fig:SIGaussTbl}
\end{figure}

That is, the smaller the $c$, the smaller is the maximal attitude change (see also Supplementary Figure~\ref{fig:SIGaussTbl} \textit{\textbf{a}}, \textit{\textbf{d}} and \textit{\textbf{g}}). For example, in Figure~S\ref{fig:SIGaussTbl} \textit{\textbf{a}}, in case the "(original) attitude towards the concept under modification", $a_i^t$, recorded on the horizontal ($x$) axis, is zero, the maximal change $M$ is $1$ (recorded on the vertical ($y$) axis). But in case $a_i^t = abs(0.5)$ then $M \approx 0.67$, if $a_i^t = abs(0.75)$ then $M \approx 0.4$, and finally, in case $a_i^t \rightarrow abs(1)$ then $M \rightarrow 0$.

Figure~S\ref{fig:SIGaussTbl} shows the simulation results for 3 different values of $c$: $c=0.8$ (top row), $c=0.5$ (middle row), and $c=0.3$ (bottom row). The two main observations are that:\\
(1) as the exposure to the news intensifies (marked by the iteration number $t$ on sub-figures \textit{\textbf{b}}, \textit{\textbf{e}} and \textit{\textbf{h}}), the attitude values tend towards the two extremities, + and -1 ( -- which is the main claim of the manuscript).\\
(2) The smaller is the $c$ is (that is, the more pronounced the hierarchical effect), the more are the "centralist" (near-zero) attitudes in case of moderate exposure (small $t$ values).\\

\end{document}